\documentclass[useAMS,usenatbib]{mn2e}
\usepackage{rotating}

 \usepackage{times}

\def\pcm3{{\rm\thinspace cm^{-3}}}

\def\contcaption{\@conttrue\SFB@caption\@captype}

\def\n_h{{\rm n_{H}}}

\def\NH1{{$N_{\rm HI}~$}}

\def\ga{{\rm\thinspace gauss}}

\def\approxlt{\mathrel{\hbox{\rlap{\lower .5ex \hbox {$\sim$}}
        \raise .15 ex \hbox{$<$}}}}
\def\approxgt{\mathrel{\hbox{\rlap{\lower .5ex \hbox {$\sim$}}
        \raise .15 ex \hbox{$>$}}}}

\def\la{\mathrel{\hbox{\rlap{\hbox{\lower4pt\hbox{$\sim$}}}\hbox{$<$}}}}
\def\ga{\mathrel{\hbox{\rlap{\hbox{\lower4pt\hbox{$\sim$}}}\hbox{$>$}}}}

\newbox\grsign \setbox\grsign=\hbox{$>$} \newdimen\grdimen
\grdimen=\ht\grsign
\newbox\simlessbox \newbox\simgreatbox \newbox\simpropbox
\setbox\simgreatbox=\hbox{\raise.5ex\hbox{$>$}\llap
     {\lower.5ex\hbox{$\sim$}}}\ht1=\grdimen\dp1=0pt
\setbox\simlessbox=\hbox{\raise.5ex\hbox{$<$}\llap
     {\lower.5ex\hbox{$\sim$}}}\ht2=\grdimen\dp2=0pt
\setbox\simpropbox=\hbox{\raise.5ex\hbox{$\propto$}\llap
     {\lower.5ex\hbox{$\sim$}}}\ht2=\grdimen\dp2=0pt
\def\simgreat{\mathrel{\copy\simgreatbox}}
\def\simless{\mathrel{\copy\simlessbox}}

\title[]{New Praesepe white dwarfs and the initial mass-final mass relation}

\author[P. D. Dobbie et al.]{P. D. Dobbie$^{1}$\thanks{E-mail:
pdd@star.le.ac.uk}  R. Napiwotzki$^{2}$ M. R. Burleigh$^{1}$ M. A. Barstow$^{1}$ D. D. Boyce$^{1}$\newauthor  S. L. Casewell$^{1}$ R. F. Jameson$^{1}$ I. Hubeny$^{3}$ G. Fontaine$^{4}$\\
$^{1}$Department of Physics and Astronomy, University of Leicester, University Road, Leicester LE1 7RH, UK\\
$^{2}$Science \& Technology Research Institute, University of Hertfordshire,
College Lane, Hatfield, AL10 9AB \\
$^{3}$Steward Observatory and Department of Astronomy, University of Arizona, Tucson, AZ 85721, USA\\
$^{4}$D\'epartement de Physique, Universit\'e de Montr\'eal, C.P.6128, Succ. Centre-Ville, Montr\'eal, Qu\'ebec, Canada H3C 3J7\\
}
\begin{document}

\date{Accepted 1988 December 15. Received 1988 December 14; in original form 1988 October 11}

\pagerange{\pageref{firstpage}--\pageref{lastpage}} \pubyear{2002}

\maketitle

\label{firstpage}

\begin{abstract}

We report the spectroscopic confirmation of four further white dwarf members of Praesepe. This 
brings the total number of confirmed white dwarf members to eleven making this the second largest 
collection of these objects in an open cluster identified to date. This number is consistent with 
the high mass end of the initial mass function of Praesepe being Salpeter in form. Furthermore, it
suggests that the bulk of Praesepe white dwarfs did not gain a substantial recoil kick velocity
from possible asymmetries in their loss of mass during the asymptotic giant branch phase of evolution.
By comparing our estimates of the effective temperatures and the surface gravities of WD0833+194,
WD0840+190, WD0840+205 and WD0843+184 to modern theoretical evolutionary tracks we have derived 
their masses to be in the range 0.72$-$0.76M$_{\odot}$ and their cooling ages $\sim$300Myrs. For an 
assumed cluster age of 625$\pm$50Myrs the infered progenitor masses are between 3.3$-$3.5M$_{\odot}$. 
Examining these new data in the context of the initial mass-final mass relation we find that it can be
adequately represented by a linear function (a$_{0}$=0.289$\pm$0.051, a$_{1}$=0.133$\pm$0.015) over 
the initial mass range 2.7M$_{\odot}$ to 6M$_{\odot}$. Assuming an extrapolation of this relation to 
larger initial masses is valid and adopting a maximum white dwarf mass of 1.3M$_{\odot}$, our results 
support a minimum mass for core-collapse supernovae progenitors in the range $\sim$6.8-8.6M$_{\odot}$.

\end{abstract}


\section{Introduction}

The initial mass-final mass relation (IFMR) characterises the amount of material stars with 
primordial masses M$\simless$10M$_{\odot}$ cast out into interstellar space during post main
sequence evolution, en route to becoming white dwarfs. Accordingly, the form of this relation is
of considerable importance to investigations relating to the chemical evolution of the Milky Way 
and galaxies in general. The details of the upper end of the IFMR also have relevance to 
supernovae studies since theoretical predictions of the rate of these explosions 
are rather sensitive to the assumed minimum mass of a core-collapse progenitor. Furthermore,
the form of the IFMR conveys information on the mass loss processes which occur during the final 
stages of stellar evolution, which are difficult to model in a physical context.
 
Arguably the most robust method with which to place (semi-) empirical constraints on the form 
of the IFMR is via the study of white dwarfs in open clusters (e.g. Weidemann 1977, Romanischin
\& Angel 1980, Weidemann 2000). Since the constituents of an open cluster have a common age, 
determinable from the main sequence turn-off mass (e.g. King \& Schuler 2005), the lifetime of
a progenitor star can be estimated from the difference between this  age and the cooling time 
of the resulting white dwarf. Subsequently, the progenitor masses can be determined by comparing
their estimated lifetimes to the predictions of stellar evolutionary models.
 
\begin{figure*}
\vspace{310pt}
\includegraphics{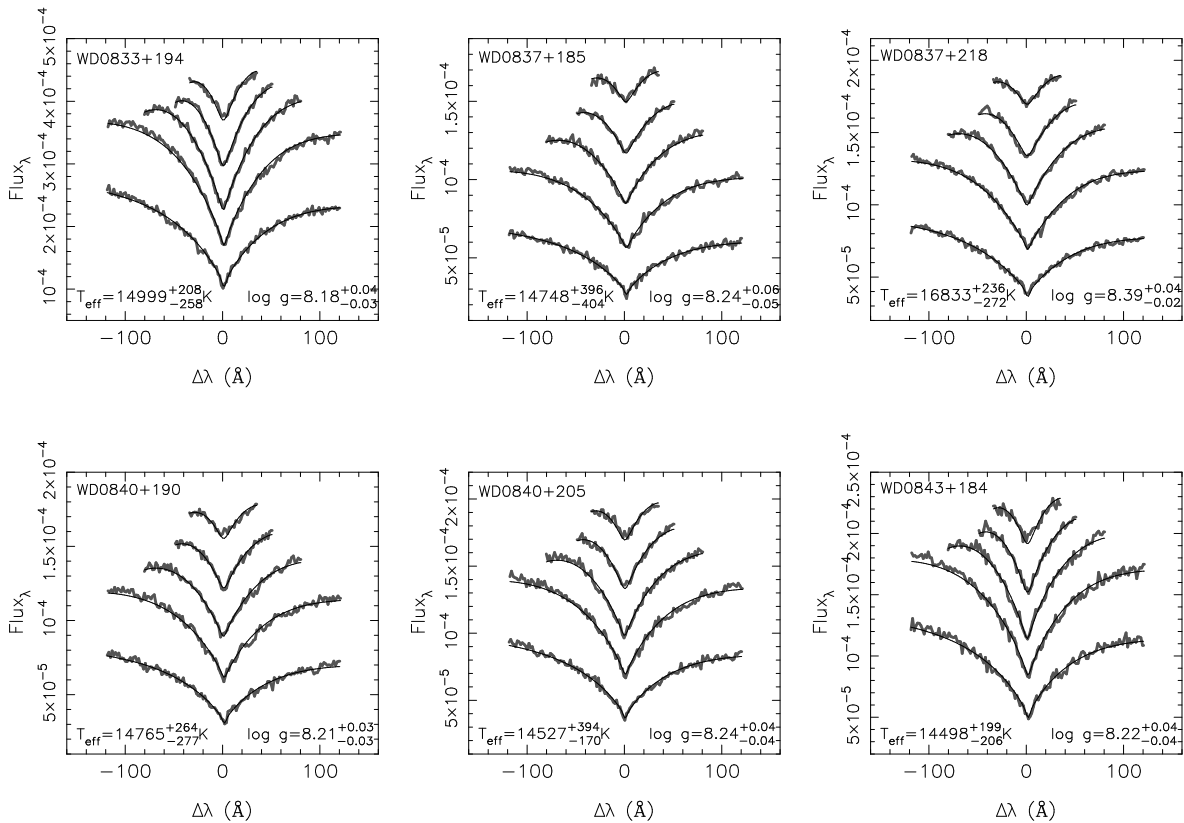}
\caption{The results of our fitting of synthetic profiles (thin black lines) to the observed Balmer
 lines, H-$\beta$ to H-8, (thick grey lines). The flux$_{\lambda}$ units are arbitrary.}
\end{figure*}

Unfortunately, until quite recently, the small numbers of WDs recovered in each open cluster ($\sim$30 
WDs in 14 clusters), their intrinsic faintness and the significant distances involved has meant 
that the uncertainties in clusters ages and in white dwarf mass determinations (and hence cooling 
time estimates), have conspired to produce significant scatter in the semi-empirical IFMR (e.g. Claver et 
al. 2001). Encouragingly, with time on 8/10m class telescopes now readily available, this situation is 
beginning to improve. For example, Kalirai et al. (2005) and Williams et al. (2005) have 
spectroscopically identified 16 and 6 likely white dwarf members of the rich but relatively distant 
open clusters M37 and M35 respectively. These two populations of white dwarfs are found to be 
consistent with a monotonically increasing IFMR for progenitor mass ranges M$\sim$2.8-3.4M$_{\odot}$ 
and M$\sim$4.5-5.5M$_{\odot}$, respectively. 

Nevertheless, while significant additional pieces of the jigsaw puzzle have recently been put in place, much 
work remains to be done before we are able to claim a thorough understanding of the IFMR.
Accordingly, we recently embarked on a search for white dwarfs members of the modestly rich and relatively
nearby Praesepe open cluster (M44, NGC2632). Using proper motions measurements and photographic photometry drawn from 
the USNO-B1.0 catalogue and SuperCosmos archive we identified 6 new candidates in a $5^{\circ} \times 
5^{\circ}$ region centered on the cluster, to add to the five previously known Praesepe white dwarfs. Futhermore, 
two of these new candidates were confirmed spectroscopically. Full details of these objects are given in 
Table 1 of Dobbie et al. (2004).

In this work we present new low resolution optical spectroscopy which confirms that the four 
remaining candidates are also white dwarfs. Additionally, we use new data to provide 
refined estimates of the effective temperatures and surface gravities of LB5959 and WD0837+218.  
We re-examine the membership status of each of these objects and present a new astrometric measurement 
of the high mass ``Praesepe''
white dwarf LB5893 which establishes that it has a proper motion which is consistent with cluster 
membership. We discuss the implications of our findings in relation to the initial mass function of 
Praesepe and kinematic effects of mass loss in the final stages of the evolution of the white dwarf 
progenitors. Adopting an age of 625$\pm50$Myrs for the cluster, we estimate the masses of these 
progenitor stars and examine our results in the context of previous work on the initial mass-final
mass relation. 

\vspace{0.1cm}

\section[]{Optical spectroscopy of the remaining four candidate white dwarf members}

We have acquired optical spectra (3200-6000\AA) of the four remaining candidate white dwarf members of Praesepe 
from Dobbie et al. (2004), using the William Herschel Telescope and the ISIS spectrograph. The 
observations were conducted during the service time nights of 2005/01/10 and 2005/11/07 and the visitor
night of 2006/02/02. Sky conditions were good on the first and third nights with clear skies and seeing 
$\sim$0.7-0.9''. On the second night there was some thin and patchy cirrus and seeing was slightly poorer,
$\sim$1.0-1.2''. All 
data were obtained on the blue arm of ISIS, using the EEV12 detector, the R300B grating and a slit 
width of 1'' to provide a spectral resolution of $\approx3.5$\AA. The total exposure time for each object
was 45 minutes made up from a series of 15 minute integrations. The CCD 
frames were debiased and flat fielded using the IRAF procedure CCDPROC. Cosmic ray hits were removed using 
the routine LACOS SPEC (van Dokkum 2001). Subsequently the spectra were extracted using the APEXTRACT package and 
wavelength calibrated by comparison with the CuAr+CuNe arc spectra. To facilitate the removal of 
the instrument signature from the science spectra we obtained observations of the spectral standard 
stars PG0843+546 (Massey et al. 1988) and G191-B2B (Oke 1990). 

\section{Analysis of the data} 

\subsection{Model white dwarf spectra}

\begin{table*}
\begin{minipage}{100mm}
\begin{center}
\caption{Details of the four new confirmed white dwarf candidate cluster members 
(top) and the two white dwarfs members identified by Dobbie et al. (2004; bottom) for which we have obtained
improved spectroscopic data. Masses and cooling times for each star have been estimated using the mixed CO core 
composition ``thick H-layer'' evolutionary calculations of the Montreal Group (e.g. Fontaine, Brassard \& 
Bergeron 2001).} 

\label{wdmass}
\begin{tabular}{lccccccc}
\hline
WD   & ID in D04 &T$_{\rm eff}$(K) & log g & M(M$_{\odot}$)  &  $\tau_{c}$(Myrs)\\
\hline

0833+194 & candidate 1 & $14999^{+208}_{-258}$ & $8.18^{+0.04}_{-0.03}$ & $0.72\pm0.02$  & $262^{+15}_{-14}$ \\ \\
0840+190 & candidate 4 & $14765^{+264}_{-277}$ & $8.21^{+0.03}_{-0.03}$ & $0.74\pm0.02$  & $288^{+14}_{-13}$\\ \\
0840+205 & candidate 5 & $14527^{+394}_{-170}$ & $8.24^{+0.04}_{-0.04}$ & $0.76\pm0.03$  & $316^{+21}_{-19}$\\ \\
0843+184 & candidate 6 & $14498^{+199}_{-206}$ & $8.22^{+0.04}_{-0.04}$ & $0.75\pm0.03$  & $308^{+20}_{-19}$\\ \\
\hline
0837+185 & candidate 2 & $14748^{+396}_{-404}$ & $8.24^{+0.06}_{-0.05}$ & $0.76\pm0.04$  & $303^{+28}_{-25}$\\ \\
0837+218 & candidate 3 & $16833^{+236}_{-272}$ & $8.39^{+0.04}_{-0.02}$ & $0.86\pm0.02$  & $267^{+14}_{-13}$\\
\hline
\end{tabular}
\end{center}
D04: Dobbie et al. (2004)
\end{minipage}
\end{table*}

The broad hydrogen Balmer lines evident in the data presented in Figure 1 are consistent with 
all four objects without prior spectroscopic data being DA white dwarfs. We have used the latest
versions of the plane-parallel, hydrostatic, non-local thermodynamic equilibrium (non-LTE) atmosphere
and spectral synthesis 
codes TLUSTY (v200; Hubeny 1988, Hubeny \& Lanz 1995) and SYNSPEC (v48; Hubeny, I. and Lanz, 
T. 2001, http://nova.astro.umd.edu/) to generate a new grid of pure-H synthetic 
spectra covering the T$_{\rm eff}$ and surface gravity ranges 10000-34000K and log g=7.0-9.0 
respectively. We have employed a model H atom incorporating the 8 lowest energy 
levels and one superlevel extending from n=9 to n=80, where the dissolution of the high lying 
levels was treated by means of the occupation probability formalism of Hummer \& Mihalas (1988), 
generalised to the non-LTE situation by Hubeny, Hummer \& Lanz (1994). All calculations included 
the bound-free and free-free opacities of the H$^{-}$ ion and incorporated a full treatment for 
the blanketing effects of HI lines and the Lyman $-\alpha$, $-\beta$ and $-\gamma$ satellite 
opacities as computed by N. Allard (e.g. Allard et al. 2004). In contrast to the grid of models 
used in our previous work where radiative equilibrium was assumed (Dobbie et al. 2004), these 
latest calculations include, where appropriate, a treatment for convective energy transport 
according to the ML2 prescription of Bergeron et al. (1992), adopting a mixing length paramater,
$\alpha$=0.6. 
During the calculation of the model structure the hydrogen line broadening was addressed in the 
following manner: the broadening by heavy perturbers (protons and hydrogen atoms) and electrons 
was treated using Allard's data (including the quasi-molecular opacity) and an approximate 
Stark profile (Hubeny, Hummer \& Lanz 1994) respectively. In the spectral 
synthesis step detailed profiles for the Balmer lines were calculated from the Stark broadening 
tables of Lemke (1997).

\subsection{Determination of effective temperatures and surface gravities}

As discussed in Dobbie et al. (2004), comparison between the models and the data is undertaken 
using the spectral fitting program XSPEC (Shafer et al. 1991). XSPEC works by folding a model 
through the instrument response before comparing the result to the data by means of a 
$\chi^{2}-$statistic. The best fit model representation of the data is found by incrementing free
grid parameters in small steps, linearly interpolating between points in the grid, until the 
value of $\chi^{2}$ is minimised. Errors in the T$_{\rm eff}$s and log g s are calculated  by stepping the 
parameter in question away from its optimum value and redetermining minimum $\chi^{2}$ until 
the difference between this and the true minimum $\chi^{2}$ corresponds to $1\sigma$ for a given 
number of free model parameters (e.g. Lampton et al. 1976). 

Given the probable age of the Praesepe cluster ($\sim$600-700Myrs), a number of the white dwarf 
members may have effective temperatures approaching that at which the H-Balmer 
lines reach their maximum equivalent width (T$_{\rm eff}$$\sim$12500-13500K). Previous 
spectroscopic studies of DA white dwarfs in this temperature regime indicate that the sensitivity
to surface gravity of the equivalent widths of the lower order Balmer lines (e.g. H-$\beta$, H-$\gamma$)
is reduced here (e.g. Daou et al. 1990). Therefore, in the present analysis all lines from H-$\beta$ to 
H-8 are included in the fitting process. Furthermore, given that we assign each line an independent 
normalisation parameter, there is the potential for obtaining two solutions for T$_{\rm eff}$ in this 
regime (e.g. Gianninas et al. 2005). Hence minimum $\chi^{2}$ has been approached 
from both the high and low temperature ends of the model grid. 

No convincing fit with T$_{\rm eff}$$\le$13500K is found to any of the datasets; the overall spectral
shape of the data (ie. lines and continuum) are well matched by models at the temperature and surface 
gravity solutions determined here. The results of our fitting procedure are given in Table 1 
and shown overplotted on the data in Figure 1. As a check for potential systematic errors (e.g. Napiwotzki 
et al. 1999) we have validated the results with an independent analysis technique (FITSB2; Napiwotzki et 
al. 2004) and a different set of model atmospheres (the LTE atmospheres used by Koester et al. 2001). 
Differences between individual results were found to be well within the error limits quoted in Table 1 and no 
significant systematic discrepancies were apparent. Nevertheless, it should be noted that the parameter 
errors quoted here are formal $1\sigma$ fit errors and may underestimate the true uncertainties.

 \begin{figure}
\vspace{260pt}
\includegraphics{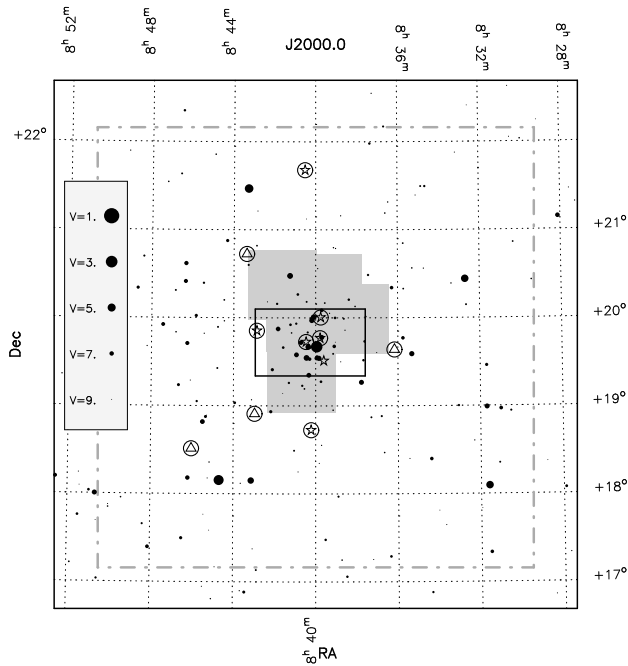}
\caption{A schematic plot of the Praesepe cluster showing stars down to V $\approx9$ and the 
areas surveyed by Anthony-Twarog (1982, 1984; solid outline) and Claver et al. (2001; grey shading). The region 
included our investigation is outlined (dashed grey line). All objects listed in Table 1 of 
Dobbie et al. (2004; open circles) and the five 'original' white dwarf cluster members (open 
stars) are also overplotted. The locations of the four new spectroscopically confirmed white dwarf 
candidate members are highlighted (open triangles).}
\end{figure}

\section{Discussion}

\subsection{White dwarf membership status, the initial mass function of Praesepe and recoil kicks}

We have used modern evolutionary tracks supplied by the Montreal group (e.g. Fontaine, Brassard 
\& Bergeron 2001) to determine the masses and cooling times of our four new white dwarfs. We have 
adopted the calculations which include a mixed CO core and thick H surface layer structure, which 
make this work consistent with other recent studies in this area (e.g. Liebert et al. 2005a).  
The masses and cooling times shown in Table 1 have been derived using cubic splines to interpolate between 
points in this grid. At the present level of precision these mass determinations are
not sensitive to our choice of core composition. However, if instead we had adopted thin H-layer 
models these estimates would be systematically lower by 0.02M$_{\odot}$. Nevertheless, we can 
conclude that all four white dwarfs have masses which are significantly larger than those 
typical of field white dwarfs, the distribution of which is found to be strongly peaked at 
0.565M$_{\odot}$ (e.g. Liebert et al. 2005b, Bergeron, Liebert \& Fulbright 1995, Marsh et al. 1997). These 
comparatively high masses and the projected spatial distribution of the white dwarfs (Figure 2) argue 
that our objects are associated with the Praesepe open cluster. 

In an effort to quantify the level of contamination by field white dwarfs in our assembly of probable
Praesepe members, we have repeated our original survey procedure on four 5$^{\circ}$$\times$5$^{\circ}$ 
fields flanking the cluster to the NE, NW, SE and SW. Only 2 objects in total have been flagged as 
candidate white dwarf members of Praesepe according to the selection criteria we applied in Dobbie et 
al. (2004). There is no guarantee that spectroscopic data would confirm either of these two objects to
be a white dwarf. Thus we conclude that contamination by field white dwarfs is at a very low level.

Nevertheless, it is worth recalling here that LB5893, one of the five ``original'' white dwarf members 
of Praesepe, was not recovered by our survey. Since Luyten (1966) measured a proper motion of $\mu_{\alpha}$cos 
$\delta$=-34 mas yr$^{-1}$, $\mu_{\delta}$=-14 mas yr$^{-1}$, albeit with large uncertainties, in 
Dobbie et al. (2004) we concluded that the USNO-B1.0 astrometry of $\mu_{\alpha}$cos $\delta$=$-$56 
mas yr$^{-1}$, $\mu_{\delta}$=$-$14 mas yr$^{-1}$, may have been adversely affected by the proximity of
the star KW195. Claver et al. (2001) previously reached a similar conclusion on finding that their 
proper motion measurement of this object, which relied on POSS-I plates for epoch 1, also failed to
confirm cluster membership.  
Given the lingering uncertainties in the proper motion of LB5893 and the rather peculiar location of
this object in the semi-empirical IFMR (see Figure 11 of Claver et al. 2001), we have used the POSS-II 
F band image (1989/11/08) and a V band image (2001/02/04) obtained from the Canada-France-Hawaii Telescope 
archive at the Canadian Astrophysical Data Center to obtain a 
refined astrometric measurement for LB5893. Our estimate of $\mu_{\alpha}$cos $\delta$=$-$34$\pm$9 
mas yr$^{-1}$, $\mu_{\delta}$=$-$18$\pm$12 mas yr$^{-1}$ supports the original proper motion determination 
of Luyten (1966) and argues strongly that LB5893 is a member of Praesepe. Therefore in the subsequent 
discussion we accept that the number of Praesepe white dwarfs is at least eleven. 

\begin{table}
\begin{minipage}{80mm}
\begin{center}
\caption{Progenitor lifetimes and corresponding masses based on the white dwarf cooling 
times shown in Table 1, the Z=0.019 stellar evolutionary models of Girardi et al. (2000) and an
assumed cluster age of 625$\pm$50Myrs.}
\label{pmass}
\begin{tabular}{lcccc}
\hline
 Progenitor & {$\tau_{\rm prog}$} & {M$_{\rm prog}$} \\
 of WD & (Myrs) & (M$_{\odot}$)  &  \\ 

\hline

0833+194 &  $363^{+52}_{-52}$ & $3.30^{+0.24}_{-0.18}$ \\ \\
0840+190 &  $337^{+52}_{-52}$ & $3.38^{+0.27}_{-0.20}$ \\ \\
0840+205 &  $309^{+53}_{-54}$ & $3.49^{+0.35}_{-0.24}$ \\ \\
0843+184 &  $317^{+53}_{-54}$ & $3.46^{+0.33}_{-0.23}$ \\ \\

\hline
0837+185 &  $322^{+56}_{-57}$ & $3.44^{+0.36}_{-0.25}$ \\ \\
0837+218 &  $358^{+52}_{-52}$ & $3.31^{+0.24}_{-0.18}$ \\ 
\hline
\end{tabular}
\end{center}
\end{minipage}
\end{table}

\begin{figure*}
\vspace{310pt}
\includegraphics{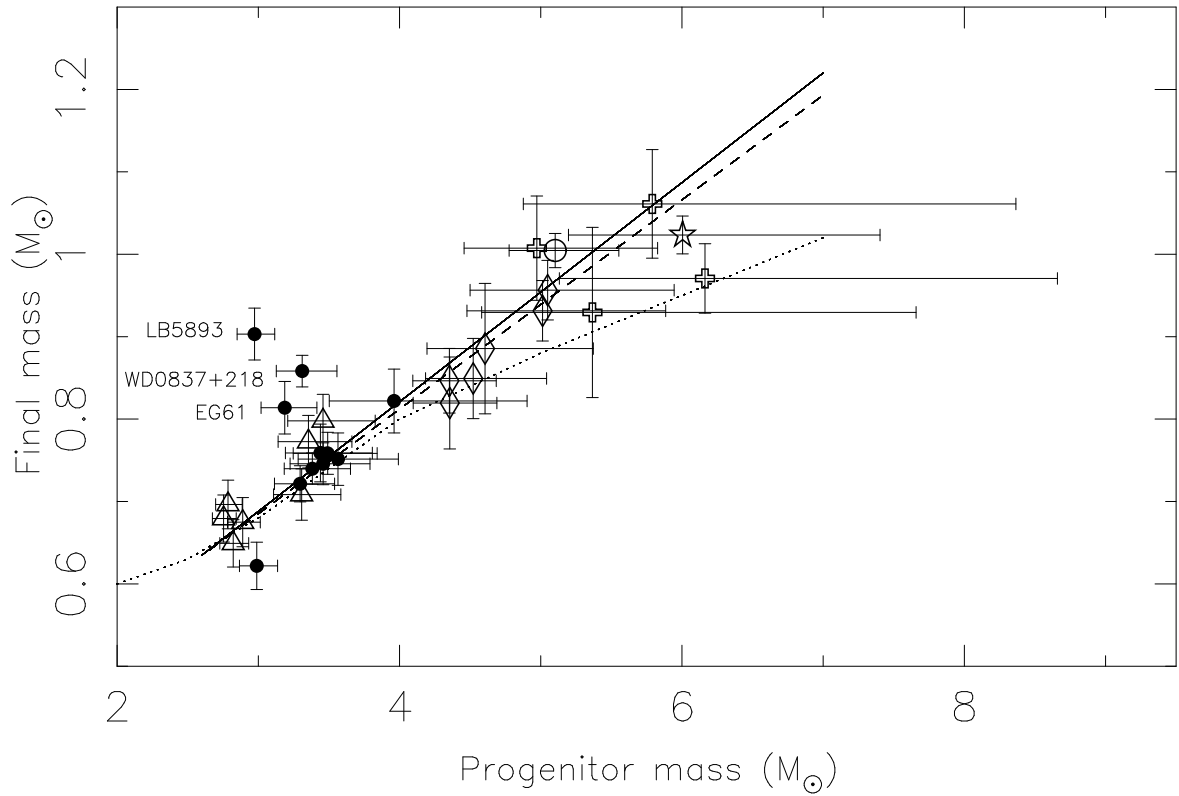}
\caption{Initial mass-final mass for the white dwarf members of the Hyades (open triangles), 
Praesepe (black circles), M35 (open diamonds), NGC2516 (open '+'s) and the Pleiades (open star). 
Sirius B is also shown (open circle). A linear fit to the data based on CO (solid line)/C(dashed line) cores 
and the relation of Weidemann (2000; dotted line) are overplotted.}
\end{figure*}

As a consequence of the optical (O-E$\le$0) and near-infrared selection criteria (no 2MASS detection or blue) 
our survey is biased against the detection of white dwarfs in unresolved binary systems with stars with
masses M$\simgreat$0.1M$_{\odot}$ (spectral types earlier than M6-7). Thus in the context of the cluster 
simulations of Williams et al. (2004), in reflection, it seems more appropriate to compare our results to 
the predicted observable number of single white dwarfs (ie. the predicted total observed number - the 
predicted number observed in binaries). We find that the observed number is inconsistent with these predictions,
for any reasonable value of the maximum mass of a white dwarf progenitor (M$_{\rm crit}$), if a steep 
powerlaw ($\Gamma$=2) shape is assumed to describe the initial mass function (IMF) of Praesepe. In constrast, 
we find that the observed number is consistent with the results of the simulations if instead the IMF is 
assumed to have been Salpeter in form for any reasonable value for M$_{\rm crit}$ (P$\approx$0.19 for 
M$_{\rm crit}$=10M$_{\odot}$).
Accordingly, we adjudge that there is not a deficit of white dwarfs, at least single objects, in this 
open cluster. If our conclusion is correct then these results suggest that the bulk of Praesepe white dwarfs 
did not gain a significant recoil kick velocity from possible asymmetries in their loss of mass during the
asymptotic giant branch phase of evolution. Fellhauer et al. (2003) find that if the mean kick velocity 
extended to a white dwarf during this phase is greater than twice the cluster velocity dispersion (the
one-dimensional velocity dispersion of Praesepe is 0.67$\pm$0.23kms$^{-1}$; Madsen et al. 2002) then a 
significant fraction of these objects evaporate from the cluster ($\simgreat$60\% at 600-700Myrs).
 
\subsection{The masses of the progenitor stars of the Praesepe white dwarfs}

The metalicities of the Praesepe and the Hyades open clusters are found, within uncertainties, to be very 
similar. For example, Cayrel de Strobel (1990) and Boesgaard \& Budge (1988) determine [Fe/H]=0.10$\pm$0.06 
and [Fe/H]=0.13$\pm$0.07 respectively for the former, while Cayrel, Cayrel de Strobel \& Campbell (1985) 
and Perryman et al. (1998) find [Fe/H]=0.12$\pm$0.03 and [Fe/H]=0.14$\pm$0.05 respectively, for the latter.
Furthermore, it has been recognised for decades that the space motions of the two clusters are comparable 
(e.g. Schwarzchild \& Hertzsprung 1913, Eggen 1992, Madsen et al. 2002), leading to the suggestion that 
Praesepe is a member of a Hyades supercluster. If this is the case it is likely that the ages of the two 
clusters are similar. 

The bulk of determinations place the age of the Hyades cluster in the range 500-900Myrs (e.g. Barry et al. 
1981, Kroupa 1995). Perryman et al. (1998) have estimated an age of 625$\pm$50Myrs by fitting theoretical 
isochrones, which included a treatment for convective overshoot, to the Hipparcos-based cluster 
Hertzsprung-Russell diagram. Claver et al. (2001) have compared modern theoretical isochrones for ages 500,
700 and 1000Myrs to the Praesepe sequence as observed in their V,V-I colour-magnitude diagram. They conclude 
that the upper main sequence is poorly reproduced by the 1000Myr isochrone but can be considered consistent
with either of the two younger models. In addition, recent work on the X-ray properties of solar type Praesepe
members reveals that these are similar to those of the F and G stars of the Hyades (Franciosini et al. 2003). 
Furthermore, we note that the most detailed study to date of the spatial distribution and dynamics of the 
cluster (Adams et al. 2002) finds no evidence for the existence of a sub-cluster as suggested by Holland et
al. (2001). Therefore, we follow Claver et al. (and more recently Ferrario et al. 2005) and in our subsequent
discussion assume the age of Praesepe is 625$\pm$50Myrs.

We have derived the lifetime of the progenitor star of each of the four new white dwarf members 
by subtracting the estimated cooling time, shown in Table 1, from the adopted cluster age. The 
results of this process are shown in Table 2. For consistency, we obtained new spectroscopy of 
comparable quality for WD0837+218 and WD0837+185 (Figure 1) and analysed it using our new grid of 
model atmospheres. Thus we also provide in Tables 1 and 2, revised effective temperatures, surface 
gravities, masses, cooling times and progenitor lifetimes for these two objects. 
We note that the bulk of the discrepancy between the new and old parameter estimates for these two 
objects stems from the different datasets used in the two analyses. The spectral datasets in 
Dobbie et al. (2004) were of comparatively low quality and the true level of uncertainty in the 
associated effective temperature and surface gravity estimates appears to have been slightly underestimated 
by our formal error analysis.

Subsequently, we have used cubic splines to interpolate between the lifetimes calculated for stars of 
solar composition by Girardi et al. (2000) and have constrained the masses of these six progenitors to 
the values shown in Table 2. The quoted errors take into account the uncertainty in the cluster age. 
The locations in initial mass-final mass space of all six of these objects and the five original Praesepe
white dwarfs, where their masses, cooling times and progenitor lifetimes and progenitor masses have been
derived on the basis of the effective temperature and surface gravity measurements of Claver et al. 
(2001), are shown in Figure 3 (filled circles). We point out that five of the Praesepe white dwarfs 
virtually sit on top of each other in this plot. 

\subsection{The initial mass-final mass relation}

Since the Hyades has a comparatively robust age determination we have added to Figure 3 the seven 
single white dwarf members of this cluster (open triangles). Their initial and final masses are derived
from the effective temperature and surface gravity measurements listed in Claver et al. (2001). As there 
still appears to be considerable uncertainty as to the age of M37 (e.g. Kharchenko et al. 2005, Kalirai et al. 2005,
Twarog et al. 1997), we do not include in Figure 3 recent data from this cluster. 
However, the eighteen white dwarfs from Praesepe and the Hyades only define the initial mass-final mass 
relation for 2.7M$_{\odot}$$\simless$M$_{\rm prog}$$\simless$4M$_{\odot}$. Therefore, we have also 
added to this plot Sirius B (open circle; Liebert et al. 2005a) and the white dwarf members of three other relatively well 
characterised but much younger open clusters ($\tau$$\simless$200Myrs), the Pleiades (open star), M35 
(open diamonds; Williams et al. 2005) and NGC2516 (open '+'s; Koester \& Reimers 1996). 

Using a new high S/N spectrum obtained with the WHT and our grid of TLUSTY models we have redetermined 
the effective temperature and surface gravity of the only known Pleiades white dwarf WD0349+247 (LB1497) to be 
T$_{\rm eff}$=$32841^{+175}_{-169}$ and log g=$8.63^{+175}_{-169}$ respectively (Figure 4). This corresponds
to a mass of 1.02$\pm$0.02M$_{\odot}$, in excellent agreement with the gravitational redshift determination 
(1.02$\pm$$^{+0.04}_{-0.05}$M$_{\odot}$; Wegner et al. 1991). 
For the white dwarf members of M35 and NGC2516 we adopt the effective temperatures and surface gravities 
listed in Table 1 of  Ferrario et al. (2005). Excepting M35, where, given the result of Kalirai et al. (2003), 
we prefer an age of 160$\pm$25Myrs, the progenitor mass for each of these white dwarfs is derived assuming 
the system/cluster age and metalicity adopted by Ferrario et al. (2005). The upper limits to the progenitor 
masses of the white dwarf members of NGC2516 and the Pleiades displayed in Figure 3 have been determined by 
supplementing the model grid of Girardi with an evolutionary calculation for a 9M$_{\odot}$ star drawn from a
grid of a previous generation of the Padova models (Bressan et al. 1993). 

\begin{figure}
\vspace{162pt}
\includegraphics{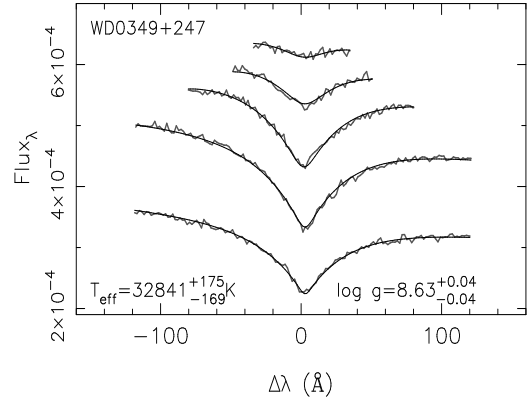}
\caption{The results of our fitting of synthetic profiles (thin black lines) to the observed Balmer
 lines of WD0349+247, H-$\beta$ to H-8, (thick grey lines). The flux$_{\lambda}$ units are arbitrary.}
\end{figure}

An examination of the thirty stars in Figure 4 indicates that the majority appear to follow rather 
closely a monotonic relation between their initial and final masses. Indeed, the white dwarf members of 
the subsolar metalicity cluster M35 ([Fe/H]$\approx$-0.3; Sung \& Bessell 1999) seem to form a natural
extension of the relation defined by the Hyades and Praesepe white dwarfs (see Figure 3). This suggests 
that any effect metalicity may have on the form of the IFMR is probably not detectable at our current level of 
precision. Nevertheless, two stars, both attributable to Praesepe, appear to deviate more significantly 
($\simgreat$3$\sigma$) from the general relation (LB5893 and WD0837+218). As briefly discussed by Claver
et al. (2001) and Dobbie et al. (2004) these white dwarfs appear to be too hot for their relatively high 
masses. We note here that their masses are coincident with the secondary peak in the mass distribution 
of white dwarfs ($>$0.8M$_{\odot}$; see Figure 13 of Liebert et al. 2005b). As discussed by Liebert et al.
it has been suggested that a significant proportion of the white dwarfs in this secondary peak may be 
the progeny of close binary systems. 

The initial mass-final mass data shown in Figure 3 can be reasonably approximated by a linear function.
In deriving the co-efficients for this we exclude the two objects which appear to deviate most strongly 
from the general relation as these most likely result from close binary evolution.  Additionally, we 
choose to reject from the our fit the magnetic white 
dwarf EG61 as the true uncertainties in the effective temperature and surface gravity determinations 
for this star, obtained from the Zeeman split profiles of the Balmer lines, could be significantly 
larger than the formal errors quoted by Claver et al. (2001). 
Thus performing a linear least squares fit to the remaining twenty-seven white dwarfs we derive fit
co-efficients a$_{0}$=0.289$\pm$0.051, a$_{1}$=0.133$\pm$0.015 (solid line). We note that 
had we adopted pure-C core white dwarf evolutionary models, the cooling times would have been
systematically larger by between 5-30Myrs, resulting in shorter progenitor lifetimes, larger progenitor
masses and a marginally flatter IMFR (a$_{0}$=0.304, a$_{1}$=0.127; dashed line). In either case our 
fit is somewhat steeper than the widely applied Weidemann (2000) relation, in particular at 
M$_{\rm prog}$$\simgreat$4M$_{\odot}$ where the latter flattens slightly (dotted line).

In standard stellar evolutionary theory, the CO core of an intermediate mass star can grow to 
$\sim$1.1M$_{\odot}$ before the carbon ignites.
The resulting degenerate NeO core collapses, 
through electron capture onto Ne, to form a neutron star. Extrapolating our fit to the initial 
mass-final mass data we find that a 1.1M$_{\odot}$ white dwarf is produced by a star with an initial mass 
of $\sim$6.1M$_{\odot}$. However, more recent theoretical modelling by Garcia-Berro et al. (1997) have shown 
that it is possible for stable more massive NeO degenerate cores to form via single star evolution. For 
example, their calculation of the evolution of a 10M$_{\odot}$ star ends in the production of a 
1.26M$_{\odot}$ NeO white dwarf. However, a 1.37M$_{\odot}$ degenerate NeO core produced in a 
similar computation for a more massive star collapses, resulting in
a type II supernova explosion. These findings led Weidemann (2000) to suggest that the upper 
limit for a white dwarf mass is $\sim$1.3M$_{\odot}$. On the basis of an extrapolation of our fit this 
corresponds to progenitor masses of 7.6(6.8-8.6)M$_{\odot}$, somewhat lower than the 10-11 M$_{\odot}$ 
suggested by the models of Garcia-Berro et al. (1997). 

There is no observational evidence to indicate that the form of the IMFR can be represented by an 
extrapolation of our linear function for M$_{\rm prog}$$\simgreat$6M$_{\odot}$. Nevertheless, if we
were to assume that such an extrapolation is valid and that the maximum white dwarf mass is 
1.3M$_{\odot}$ then these results argue that the minimum mass of 
a core collapse supernova progenitor lies in the range $\sim$6.8-8.6M$_{\odot}$. We note that Ratnatunga \& van den Bergh 
(1989) estimate a Galactic type II supernova rate of 1.1$^{+1.5}_{-0.6}$ century$^{-1}$ for a minimum
 progenitor mass of 8M$_{\odot}$. 

\section{Summary}

We have spectroscopically confirmed four more white dwarf members of Praesepe and argued that there 
are at least eleven white dwarfs in this cluster. We find that this number is consistent with that 
expected if the initial mass function of Praesepe was Salpeter in form. We suggest that this consistency
indicates that few white dwarfs have evaporated from the cluster since their formation. If this is 
correct then any recoil velocity kick arising from possible asymmetries in mass loss during the final 
stages of stellar evolution is likely smaller than $\sim$2.5kms$^{-1}$. We find most stars appear to 
follow, relatively closely, a monotonic relation between their initial and final masses. This relation 
can be reasonably approximated by a linear function with coefficients a$_{0}$=0.289$\pm$0.051,
a$_{1}$=0.133$\pm$0.015. Extrapolation of this linear function beyond initial masses $\sim$6M$_{\odot}$
supports a minimum mass for a type II supernova progenitor in the range $\sim$6.8-8.6M$_{\odot}$ for an assumed 
maximum white dwarf mass of 1.3M$_{\odot}$.

\section*{Acknowledgments}
PDD is a PPARC postdoctoral research associate. RN and MBU acknowledge the
support of PPARC Advanced Fellowships. DDB and SLC acknowledge receipt of 
PPARC studentships. The WHT is operated on the island of
La Palma by the Isaac Newton Group in the Spanish Observatorio del Roque de los
Muchachos of the Instituto de Astrofisica de Canarias. We thank Pierre Leisy and 
Javier Licandro for conducting some of these observations as part of the ING 
service time programme. We express our gratitude to Jay Holberg for a helpful 
report on this manuscript.

\appendix

\bsp

\label{lastpage}

\end{document}